\documentclass[reprint,aps,prl]{revtex4-1}

\usepackage{graphicx}
\usepackage{epsfig}
\usepackage{epstopdf}
\usepackage{subfigure,color}
\usepackage{dcolumn}
\usepackage{bm}
\usepackage{amsmath,amssymb}
\usepackage{amsfonts}
\usepackage{esvect}
\usepackage{harpoon}
\usepackage{hyperref}
\usepackage{amsmath,accents}
\usepackage{verbatim}
\usepackage[dvipsnames]{xcolor}

\def\.{\cdot}

\def\##1{{\bf #1\mit}}
\def\_#1{{\bf #1\mit}}
\def\-#1{{\bf #1\mit}}
\def\=#1{\overline{\overline #1}}

\begin{document}

	\newcommand{\red}[1]{\textcolor{red}{#1}}
 \newcommand{\blue}[1]{\textcolor{blue}{#1}}
  \newcommand{\green}[1]{\textcolor{SeaGreen}{#1}}
 \title{Expanding momentum bandgaps in photonic time crystals through resonances}

 \author{X.~Wang$^{1,2,6}$}
 \email{Corresponding author: xuchen.wang@hrbeu.edu.cn}
	\author{P.~Garg$^{3,6}$}
 \email{Corresponding author: puneet.garg@kit.edu}
	\author{M.~S.~Mirmoosa$^4$}
 \author{A.~G.~Lamprianidis$^3$}
	\author{C.~Rockstuhl$^{2,3}$}
 \author{V.~S.~Asadchy$^5$}
	\affiliation{$^1$\mbox{Qingdao Innovation and Development Base, Harbin Engineering University, Qingdao 266400, China}\\
 $^2$\mbox{Institute of Nanotechnology, Karlsruhe Institute of Technology, Karlsruhe, Germany}\\
 $^3$\mbox{Institute of Theoretical Solid State Physics, Karlsruhe Institute of Technology, Karlsruhe, Germany}\\
  $^4$\mbox{Department of Physics and Mathematics, University of Eastern Finland, Joensuu, Finland}\\
 $^5$\mbox{Department of Electronics and Nanoengineering, Aalto University, Espoo, Finland}\\
 $^6$\mbox{These authors contributed equally: X. Wang, P. Garg}
 }

\begin{abstract}
The realization of photonic time crystals is a major opportunity but also comes with significant challenges. The most pressing one, potentially, is the requirement for a substantial modulation strength in the material properties to create a noticeable momentum bandgap. Reaching that noticeable bandgap in optics is highly demanding with current, and possibly also future, material platforms since their modulation strength is small by tendency. Here, we demonstrate that by introducing temporal variations in a resonant material, the momentum bandgap can be drastically expanded with modulation strengths in reach with known low-loss materials and realistic laser pump powers. The resonance can emerge from an intrinsic material resonance or a suitably spatially structured material supporting a structural resonance. Our concept is validated for resonant bulk media and optical metasurfaces and paves the way toward the first experimental realizations of photonic time crystals.
	\end{abstract}
	\maketitle
Within the domain of time-varying systems \cite{galiffi2022photonics,Tretyakov2023}, an intriguing concept that has recently garnered interest is that of photonic time crystals (PTCs)~\cite{zurita2009reflection1,lustig2018topological,asgari_photonic_2024}. PTCs are artificial materials whose electromagnetic properties are uniform in space but periodically modulated in time. This temporal periodicity generates a momentum bandgap in which light exponentially grows over time. Such behavior results in exotic light-matter interaction, including amplification of spontaneous emission of an excited atom~\cite{lyubarov2022amplified}, subluminal Cherenkov radiation~\cite{dikopoltsev2022light}, superluminal momentum-gap solitons~\cite{pan2023superluminal}, and others.

Light amplification in the momentum bandgap of PTCs is a parametric process, distinct from traditional methods in nonlinear optics~\cite{khurgin2023photonic,asgari_photonic_2024}. 
In particular, while backward optical parametric amplification under transverse pumping shares similarities in geometry and medium eigenmodes~\cite{ding_transversely_1995,lanco2006semiconductor}, it is limited by an inherent finite growth of light energy due to the oscillatory nature of light propagation~\cite{khurgin2023photonic}. 
Nevertheless, each process has an optimal regime for practical use. Backward optical parametric amplification is beneficial for moderate values of parametric gain (coupling coefficient) due to the positive feedback in space. Conversely, PTCs excel when the parametric gain is sufficiently large while pump depletion is still minor~\cite{khurgin2023photonic}. Moreover, compared to the difference frequency generation processes, amplification in PTCs has advantages: it is phase-independent, eigenmodes grow temporally rather than spatially, and phase-matching is easier to achieve~\cite[Sec.~4.3]{asgari_photonic_2024}.

While PTCs were experimentally confirmed at microwave frequencies~\cite{reyes2015observation,park2022revealing,wang2023metasurface}, designing them at optical frequencies remains a prime challenge. Indeed, the material temporal modulation must be extremely fast, typically twice the oscillation frequency of the light that probes the response, and the relative change of the refractive index of the crystal $\Delta n/n$ must be comparable to unity~\cite{zurita2009reflection1,lustig2018topological,hayran2022homega}. 
This can be achieved only with all-optical modulation of the refractive index~\cite{williamson2020integrated} (e.g., via third-order optical nonlinearities). 
In particular, the nonlinear Kerr effect under the assumption of an undepleted pump is described by linear coupled-mode equations with time-varying material refractive index.
However, nonlinear effects in low-loss materials are very weak, yielding a relative change in the refractive index to saturate at less than 1\%~\cite{borchers2012saturation}. 
Recently, transparent conductive oxides exhibiting relative changes in the refractive index of the order of 100\% were suggested as alternative material candidates to synthesize PTCs~\cite{alam2016large,bohn2021spatiotemporal,zhou2020broadband,caspani2016enhanced,tirole2023double}.  
However, they demand extremely high pumping power densities up to tens of TW/${\rm cm}^3$~\cite{hayran2022homega}. When combined with significant material dissipation, this could lead to rapid thermal damage to the material. 
Therefore, only time refraction has been experimentally realized in transparent conducting oxides to date~\cite{lustig2023time}. 

To avoid the aforementioned material-related obstacles, we introduce a distinct approach to designing PTCs. 
We capitalize on artificial composites that support high-quality resonances rather than seeking new materials with improved nonlinear characteristics. 
Interestingly, a similar generic idea was recently suggested in Ref.~\cite{khurgin_energy_2023}, however, no practical design or implementation was demonstrated.  
The approach based on resonant artificial composites allows us to create PTCs with pronounced momentum bandgaps with significantly reduced required modulation strength ($\Delta n/n$ ratio) in reach with known low-loss materials and realistic laser pump powers. This provides the first material platform to realize a PTC at optical frequencies.
We validate our concept of a resonant PTC for bulk materials and optical metasurfaces, assuming realistic material losses. 

\section{Results} 
\subsection{Photonic time crystals made of intrinsically resonant media} 



In the following, we initially demonstrate that temporally modulating the resonance frequency of a bulk material with Lorentzian dispersion significantly increases the momentum bandgap size compared to modulating the corresponding plasma frequency.

For a Lorentzian material, if the plasma frequency changes in time, the temporal permittivity in the frequency domain is written as $\epsilon(\omega, t) =1+\omega_{\rm p}^2(t)/(\omega_{\rm r0}^2-\omega^2+j\gamma\omega)$~\cite[Sec.~4.3]{Fleury2022TVD}. Here, $\omega_{\rm p}(t)$, $\omega_{\rm r0}$, and $\gamma$ represent the plasma frequency, resonance frequency, and damping factor, respectively. Considering a time-harmonic modulation of the plasma frequency with the magnitude $m$ and modulation frequency $\omega_{\rm m}$, i.e., $\omega_{\rm p}^2(t) = \omega_{\rm p0}^2  [1+m \cos (\omega_{\rm m} t)]$, the above expression for the permittivity reduces to $\epsilon(\omega, t) = 1 + \chi(\omega)[1 + m \cos(\omega_{\rm m} t)]$, where $\chi(\omega)$ denotes the stationary susceptibility of the material.

\begin{figure}[t!]
\centerline{\includegraphics[width= 1\columnwidth]{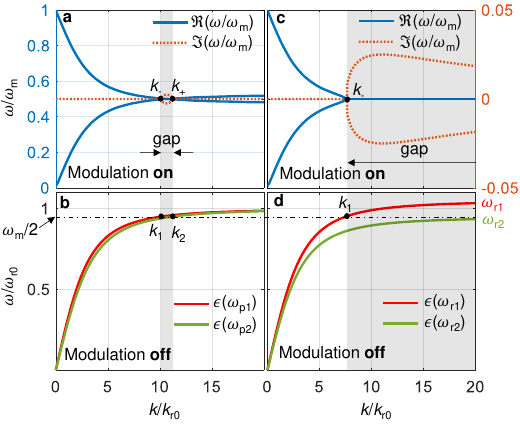}}
\caption{Band structures of time-varying Lorentzian material. (a) Band structure of a time-modulated bulk material when its plasma frequency is modulated harmonically.
(b) Dispersion relations of a stationary (not modulated) bulk material when its plasma frequency is $\omega_{\rm p1}$ (red line) and $\omega_{\rm p2}$ (green line). 
(c) Band structure of a time-modulated bulk material when its resonance frequency is modulated harmonically. 
(d) Dispersion relations of a stationary (not modulated) bulk material when its resonance frequency is $\omega_{\rm r1}$ (red line) and $\omega_{\rm r2}$ (green line).
In all the subfigures, the wavenumber is normalized by $k_{\rm r0}=\omega_{\rm r0}\sqrt{\epsilon_0\mu_0}$, and the material parameters are chosen as $m=0.2$, $\omega_{\rm m}/2=0.95\omega_{\rm r0}$, $\gamma=0$, and $\omega_{\rm p0}=3.5 \omega_{\rm r0}$. A time convention of $e^{j\omega t}$ is considered, with $j$ being the imaginary unit.
} 
\label{Fig:bulk}
\end{figure}

When solving for the eigenmodes to Maxwell's equations in this time-varying medium, the temporal modulation splits the degenerate eigenmodes at $\omega=\omega_{\rm m}/2$, resulting in a momentum bandgap (see derivations in Supplementary Section 1). 
To illustrate this phenomenon, we plot the dispersion relation in the first Brillouin zone in Fig.~\ref{Fig:bulk}(a), assuming $m=0.2$. Inside the momentum bandgap, the eigenfrequency has two complex conjugate solutions corresponding to exponentially decaying and growing modes in time.

To analytically determine the bandgap size, the weak-modulation approximation is used \cite{asadchy2022parametric} to calculate the bandgap edges ($k_-$ and $k_+$ indicated in Fig.~\ref{Fig:bulk}(a)). However, this powerful tool provides a closed-form solution for the edges and highlights a crucial observation. 
The bandgap edges $k_-$ and $k_+$ precisely correspond to the eigenwavenumbers $k_1$ and $k_2$ of two non-modulated materials with scaled susceptibilities $\chi(\omega)(1\pm m/2)$ evaluated at the frequency $\omega=\omega_{\rm{m}}/2$ (see Supplementary Section 1). 
This correspondence is general and applies to all PTC topologies with $m \ll 1$ explored in this work, including the low-loss materials. 
Accordingly, Fig.~\ref{Fig:bulk}(b) depicts the dispersion relations of the two stationary materials with those scaled susceptibilities. Since $\chi$ is linearly proportional to $\omega_{\rm p}^2$, the two materials can be alternatively described with scaled plasma frequencies $\omega_{\rm p1}^2=\omega_{\rm p0}^2(1-m/2)$ and $\omega_{\rm p2}^2=\omega_{\rm p0}^2(1+m/2)$. Thus, by plotting the dispersion relation of the material with no modulation, one can determine the momentum bandgap size for a given modulation. We emphasize this property in Figs.~\ref{Fig:bulk}(a) and (b) with the vertically shaded region. Importantly, this correspondence provides a simple visual insight into how a given material dispersion affects the bandgap size and position. It is evident from Figs.~\ref{Fig:bulk}(a) and (b) that the temporal modulation of the plasma frequency results in a relatively narrow bandgap because of the tiny variation of the dispersion curves of a stationary material with fixed resonance frequency ($k_1 \approx k_2$).

On the contrary, the situation drastically differs when the resonance frequency $\omega_{\rm r}$ varies in time.
To show this, we assume that the intrinsic resonance frequency changes harmonically so that $\omega_{\rm r}^2(t) = \omega_{\rm r0}^2[1 + m\cos(\omega_{\rm m} t)]$. Following the above-described correspondence between the stationary and modulated material scenarios, we plot in Fig.~\ref{Fig:bulk}(d) the dispersion relation of a stationary bulk material for two values of the resonance frequency: $\omega_{\rm r1}^2 = \omega_{\rm r0}^2(1 + m/2)$ and $\omega_{\rm r2}^2= \omega_{\rm r0}^2(1 - m/2)$. 
We observe that for the same modulation strength of $m=0.2$, the two dispersion curves now differ significantly at $\omega=\omega_{\rm m}/2$.
For $\omega_{\rm r2} < \omega_{\rm m}/2 < \omega_{\rm r1}$, the horizontal dashed line at $\omega=\omega_{\rm m}/2$ intersects only with one of the dispersion curves, at $k=k_1$ (Fig.~\ref{Fig:bulk}(d)), indicating a semi-infinite momentum bandgap extending from $k=k_1$ to $k=+\infty$, as shown by the gray-shaded region. The prediction from the dispersion curves of stationary materials agrees excellently with the rigorously computed band structure (see Supplementary Section 2) shown in Fig.~\ref{Fig:bulk}(c).
Moreover, the imaginary part of the dispersion as a function of the momentum (orange dashed line) sensitively depends on the modulation frequencies, which might allow us to engineer the amplification rates.

In this conceptual analysis, we disregard possible effects of the nonlocal response of the material at very high $k$-values~\cite{buddhiraju_absence_2020}, which would make the bandgap in Fig.~\ref{Fig:bulk} finite but still very large. In our suggestion in the last section, nonlocal effects are fully considered.
Furthermore, we also neglected losses in Fig.~\ref{Fig:bulk}. Possible losses cause a positive offset in the imaginary parts of eigenfrequencies, reducing the temporal amplification rates (see Supplementary Section 3). However, the bandgap size remains, and the amplification rate is still significantly higher compared to modulating the plasma frequency in the lossless case.

\begin{figure}[t!]
\centerline{\includegraphics[width= 1\columnwidth]{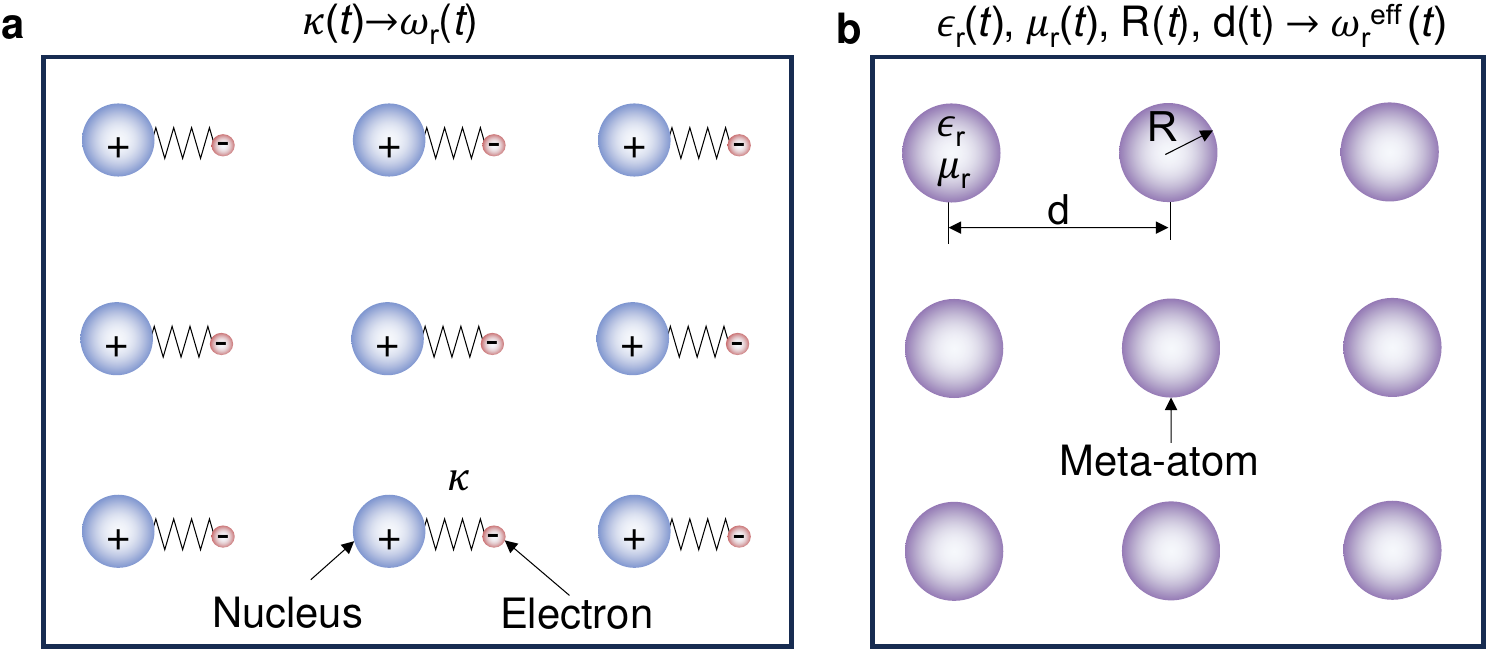}}
\caption{Conceptual realization method of time-varying medium with resonance frequency modulated. (a) Illustration of the temporal modulation of the intrinsic resonance frequency of a bulk material. The modulation can be described by a time-varying effective spring constant $\kappa(t)$. (b) Illustration of the temporal modulation of the resonance frequency of a metamaterial consisting of meta-atoms. The modulation can be achieved in various ways, including modulating the permittivity $\epsilon_{\rm r}$ or permeability $\mu_{\rm r}$ of the material from which the meta-atoms are made, the size of the meta-atoms $R$, or the metamaterial periodicity $d$. }
\label{Fig:concept}
\end{figure}

The temporal modulation of the intrinsic resonance frequency of bulk material is achieved, for example, through strong dynamic electric biasing~\cite{serra2023homogenization} but, in practice, could be very challenging. Figure~\ref{Fig:concept}(a) illustrates this scenario where an external electric field modulates the effective spring constants $\kappa$ of each nucleus-electron oscillator. However, instead of modulating the intrinsic resonance frequency of the natural atoms of a material, we propose a metamaterial concept where the resonances of spatially structured meta-atoms are modulated from which a metamaterial is assembled, as shown in Fig.~\ref{Fig:concept}(b). 
Probably, the most practical scenario is when the properties of the material from which the meta-atoms are made are modulated, which can be achieved with current optical modulation techniques~\cite{alam2016large,bohn2021spatiotemporal,zhou2020broadband,caspani2016enhanced}.



\begin{figure*}
\centerline{\includegraphics[width= 2.1\columnwidth]{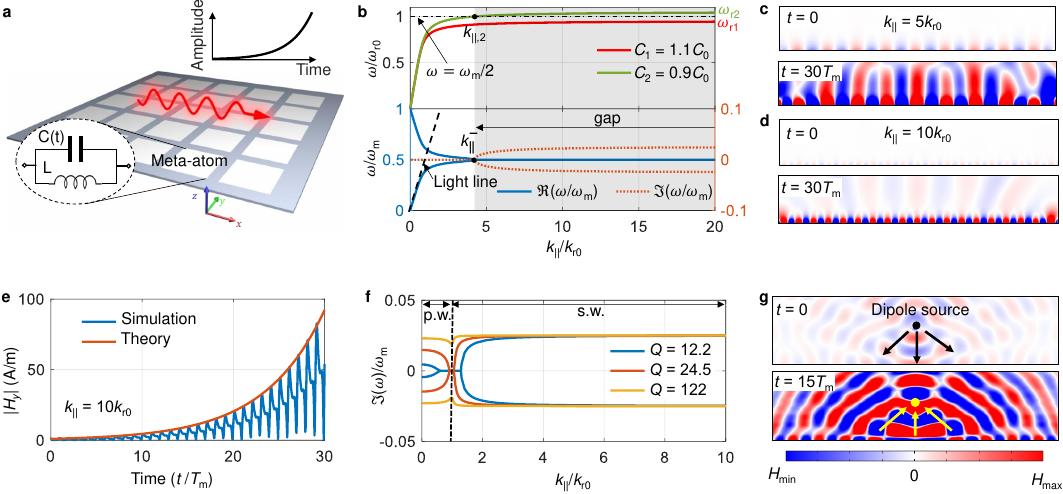}}
\caption{Analysis of time-varying $LC$ resonant metasurface. (a) Generic illustration of a time-varying resonant $LC$ metasurface. Each meta-atom is described by a time-modulated surface capacitance $C(t)$ and a constant surface inductance $L$. The red arrow indicates a surface eigenmode whose amplitude is growing in time due to its wavenumber being inside the momentum bangdap of the metasurface. 
(b) Upper: Dispersion relations of a stationary metasurface for the two scaled values of the surface capacitance. Lower: Band structure of the time-modulated resonant metasurface. The horizontal axis is normalized by $k_{\rm r0}=\omega_{\rm r0}/c$.
The gray region emphasizes the bandgap in the lower panel and the variation of eigenwavenumber at $\omega= \omega_{\rm m}/2$ in the upper panel. 
(c) and (d) magnetic field snapshots for momenta $k_{||}=5k_{\rm r0}$ and $k_{||}=10k_{\rm r0}$ above the metasurface at the time moment when modulation switches on ($t =0 $) and after some time passed ($t= 30 T_{\rm m}$ with $T_{\rm m} = 2 \pi /\omega_{\rm m}$). The complete field evolution animation is available in movie~S1. 
(e) Time evolution of the magnetic field above the metasurface ($z=0$) calculated with full-wave simulations and analytically from the band structure. The theoretical amplitude is calculated as $H_y(t)=H_{y0} \exp{[\Im(\omega)t]} $ where $H_{y0}=1$~A/m is the initial field at $t=0$ and $\Im(\omega) = 0.025\omega_{\rm m}$.
(f) Imaginary part of the eigenfrequency for metasurfaces with different quality factors. The real part of eigenfrequency is fixed as $\Re(\omega)=\omega_{\rm m}/2$. The black dashed line separates propagating waves (p.w.) and surface waves (s.w.). The quality factor of the stationary metasurface can be qualitatively described as a quality factor of an $RLC$ circuit (where $R=\sqrt{\mu_0/\varepsilon_0}$ is the free-space characteristic impedance which is connected in parallel to the metasurface equivalent circuit), that is, $Q = \omega_{\rm r0}C_0 \sqrt{\mu_0/\varepsilon_0}$~\cite[Sec.~6.1]{pozar2011microwave}. The quality factor is tuned by varying the values of $C_0$ and $L$, while keeping $\omega_{\rm r0}=1/\sqrt{LC_0}$ constant. Importantly, such a regime of the open bandgap for propagating modes is inherent to metasurfaces with high quality factors and cannot occur in non-resonant metasurfaces like those in Ref.~\cite{wang2023metasurface}. (g) The magnetic field evolution for a dipole source positioned 1.5 wavelengths above the time-varying $LC$ metasurface with $Q=122$. Due to the infinite bandgap, all momenta $k_{||}$ are amplified by the metasurface. 
 In all the subfigures, $m=0.2$ was chosen. In Figs.~\ref{Fig:bigfigure}(c), (d), and (g), the excitation source is turned off after modulation starts. }
\label{Fig:bigfigure}
\end{figure*}

%

%

\subsection{Photonic time crystals based on resonant metasurfaces}
Next, we realize the PTC in a metasurface geometry to steer our consideration to a practical scenario. The metasurface is assumed to support surface waves bounded to the $xy$-plane (Fig.~\ref{Fig:bigfigure}(a)). 
The meta-atoms are deeply sub-wavelength and all identical, resulting in a spatially homogeneous metasurface at $z=0$. Similarly to bulk PTCs that exhibit momentum bandgaps for bulk propagating eigenmodes, the metasurface-based counterparts support momentum bandgaps for surface eigenmodes~\cite{wang2023metasurface}.

A resonant impenetrable metasurface can be described by effective reactive parameters: the surface capacitance $C$ and surface inductance $L$ connected in parallel (Fig.~\ref{Fig:bigfigure}(a)) \cite{tretyakov2003analytical}. Such a generic $LC$-model describes the interaction of light with a time-modulated metasurface independent of any specific geometry and operational frequency.
At optical frequencies, as we show in the following section, such a resonant metasurface can be a two-dimensional array of spherical nanoparticles made of a material with a time-modulated dielectric constant. At microwave frequencies, the implementation can be a mushroom-type high-impedance surface with varactors embedded between adjacent patches~\cite{sievenpiper1999high,wang2023metasurface}.

In our model, the surface capacitance varies time-harmonically as $C(t)=C_0 [1+m\cos(\omega_{\rm m}t)]$, while the surface inductance $L$ is constant. 
Such a configuration effectively modulates the resonance frequency in time. 
Without loss of generality, we consider transverse-magnetic (TM) polarization for the surface waves.
Similarly to the observations in the previous section, the momentum bandgap size of such a time-varying metasurface can be estimated by the dispersion curves of two stationary LC surfaces with surface capacitance $C_1=C_0(1+\frac{m}{2})$ and  $C_2=C_0(1-\frac{m}{2})$ (see Supplementary Section~4). For verification, we plot the dispersion curves of the two stationary surfaces and the time-modulated metasurfaces for $m=0.2$ in Fig.~\ref{Fig:bigfigure}(b). When modulating at $\omega_{\rm m}=2\omega_{\rm r0}$, the momentum bandgap has only one edge, and the size extends from $k_{||} \approx 4k_{\rm r0}$ to infinity, in the absence of nonlocal effects. We stress that akin to the bulk scenario, if realistic material losses are included, the bandgap size remains the same (see Supplementary Section 4). 

Figures~\ref{Fig:bigfigure}(c) and (d) show fields calculated from numerical simulation for modes with two distinct momenta, i.e., $k_{||}=5k_{\rm r0}$ and $k_{||}=10k_{\rm r0}$ inside the momentum bandgap. Both modes are effectively amplified after modulating for $30 T_{\rm m}$. The simulated field amplitude evolution matches very well with theoretical predictions, as illustrated in Fig.~\ref{Fig:bigfigure}(e).

As the quality factor of the metasurface increases, the size and strength of the momentum bandgap improve.
Figure~\ref{Fig:bigfigure}(f) shows that metasurfaces with a higher Q-factor provide wider momentum bandgaps for surface waves with larger amplification rates, assuming the same modulation function. In comparison, the metasurface discussed in Figs.~\ref{Fig:bigfigure}(b)--(e) has a quality factor of $Q=2.44$.
Moreover, for sufficiently large Q-factors ($Q \geq 9.75$), a second momentum bandgap opens inside the light cone, i.e., for propagating waves. 
The size of the second bandgap grows with the quality factor of the metasurface since resonances with longer lifetimes suffer from smaller radiation loss and need a weaker modulation to maintain the same amplification rate. When the quality factor takes sufficiently large values, the two bandgaps merge, and the metasurface can amplify incident waves with all possible momenta $k_{||}$ (see Fig.~\ref{Fig:bigfigure}(f)).

\begin{figure*}
\centerline{\includegraphics[width= 2.1\columnwidth,trim=1 1 0.5 1,clip]{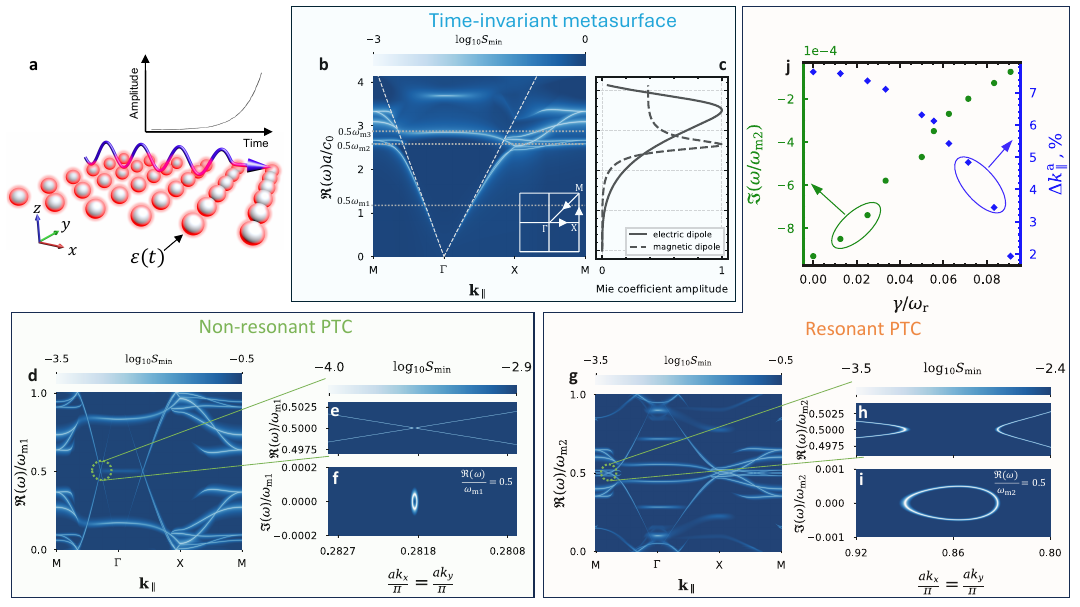}}
\caption{
Illustration of the momentum bandgap enhancement for surface waves. (a) A representative design of a PTC based on an optical time-varying resonant metasurface made of dielectric nanospheres. The nanospheres are arranged in an infinite square lattice with the period $a$. The radius of each nanosphere is $R$. The purple arrow indicates a surface eigenmode whose amplitude grows in time due to its wavenumber being inside the momentum bandgap of the metasurface. (b) Band structure of a time-invariant metasurface. The color denotes the lowest singular value $S_\mathrm{min}$ of the matrix in Eq.~\eqref{eq: Tmat_BS}. The white dashed lines represent the light lines.
(c) The two lowest-order Mie coefficients of an isolated time-invariant nanosphere. The vertical scale is the same as in (b). 
(d) Band structure of a time-varying metasurface with modulation frequency $\omega_\mathrm{m1}$ (non-resonant case). (e) Zoomed band structure in the green highlighted region of (d).  
(f) The corresponding imaginary part of the frequency for fixed $\Re(\omega)=\omega_\mathrm{m1}/2$. 
(g)--(i) Original and zoomed band structures of a time-varying metasurface with modulation frequency $\omega_\mathrm{m2}$ (resonant case). (j) Imaginary part of the eigenfrequency $\Im{(\omega)}$ at the bandgap center and the relative amplification momentum bandwidth $\Delta k^\mathrm{a}_{||}=2 |(k^\mathrm{a2}_{||}-k^\mathrm{a1}_{||}) /(k^\mathrm{a2}_{||}+k^\mathrm{a1}_{||})|$ as a function of the damping factor $\gamma$.
Here, $k^\mathrm{a1}_{||}$ and $k^\mathrm{a2}_{||}$ correspond to those values of $k_{||}$ for a bandgap where $\Im{(\omega)}=0$ (see Supplementary Fig.~S7(a)). Note that the band structures shown in (b), (d), and (g) accommodate the eigenmodes for both transverse magnetic (TM) and transverse electric (TE) polarizations. However, we optimize the metasurfaces for TE-waves in (e)--(f) and (h)--(i). Further, note that in (g), the momentum bandgap is incomplete since there exist bands at the $\Re(\omega)> \omega_\mathrm{m2}/2$ within the bandgap. Nevertheless, in contrast to spatial photonic crystals with their energy bandgaps, the modes inside an incomplete momentum bandgap are always dominant due to their amplifying nature~\cite{lustig2018topological}. 
}
\label{fig:sw}
\end{figure*}

To demonstrate this infinite momentum bandgap, we place a dipole emitter above the metasurface (see Fig.~\ref{Fig:bigfigure}(g)). The dipole radiation includes a wide spectrum of momenta, as shown in the upper panel of the figure. Once the temporal modulation of the metasurface is on, waves with all different momenta are amplified and radiated in the specular and retro-directions with respect to the source, see the lower panel in Fig.~\ref{Fig:bigfigure}(g). This leads to interesting possibilities such as amplified emission and lasing of light from a radiation source~\cite{lyubarov2022amplified}. In contrast to the idea suggested in~\cite{lyubarov2022amplified}, due to the significantly enhanced bandgap, it is possible here to amplify emission with a large and, in principle, tunable spectrum of wavenumbers. This provides opportunities for beam shaping of the amplified signal and for creating perfect lenses~\cite{pendry2000negative}. Indeed, the evanescent wave content of the source radiation can be reconstructed effectively thanks to the amplification of the wide range of $k_{||}$. In Supplementary Section 5, we demonstrate that evanescent and propagating wave components of the radiating dipole are amplified by the metasurface in reflection and transmission regimes.


%

\subsection{Optical implementation}

To provide a feasible optical realization of the resonant PTC, we consider a penetrable metasurface surrounded by air and consisting of dielectric nanospheres that are made of a material with a time-varying permittivity (see Fig.~\ref{fig:sw}(a)). Each nanosphere effectively behaves as an $LC$ resonator as it supports Mie resonances~\cite{Rahimzadegan2020light}. For simplicity, we initially ignore material dispersion. The permittivity associated with each nanosphere reads $\varepsilon(t)= 1+\chi_0[1+m\mathrm{cos}(\omega_\mathrm{m}t)]$. 
Varying the permittivity in time modulates the Mie resonance frequencies of the nanospheres (see Fig.~\ref{Fig:concept}(b)).
In the following, we rely on the T-matrix method to study the optical response from such a metasurface~\cite{garg2022modeling} (see Methods and Supplementary Section 6 for details). 

First, using Eq.~\eqref{eq: Tmat_BS} in Methods, we calculate the band structure of a time-invariant metasurface substituting $m=0$ and $\omega_\mathrm{m}=0$ (Fig.~\ref{fig:sw}(b)). One can see nearly flat bands near the $M$ and $X$ points of the spatial Brillouin zone due to the Mie resonances of the single nanosphere. For comparison, we plot the dipolar Mie coefficients of a time-invariant isolated nanosphere in Fig.~\ref{fig:sw}(c) \cite{Rahimzadegan2020light}. This figure shows magnetic and electric dipolar resonances, explaining the flat bands in Fig.~\ref{fig:sw}(b).

\begin{figure*}
\centerline{\includegraphics[width= 2.1\columnwidth,trim=4 4 4 2,clip]{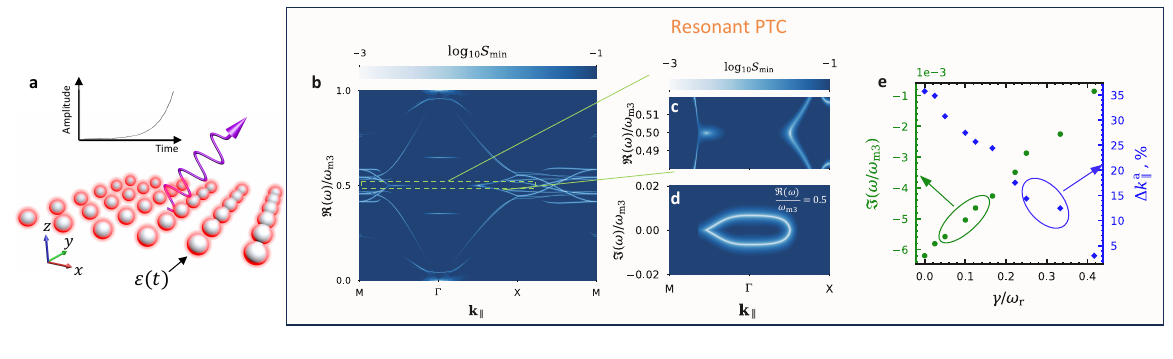}}
\caption{
Illustration of a large momentum bandgap for propagating waves. (a) Optical time-varying metasurface designed to exhibit a momentum bandgap for propagating waves. The purple arrow indicates a propagating eigenmode whose amplitude is growing in time. The geometry of the metasurface is the same as in Fig.~\ref{fig:sw}(a).
(b) Band structure of a time-varying metasurface with modulation frequency $\omega_\mathrm{m3}$. (c) Zoomed band structure in the green highlighted region in (b). (d) The corresponding imaginary part of the frequency for fixed $\Re(\omega)=\omega_\mathrm{m3}/2$. (e) Imaginary part of the eigenfrequency $\Im{(\omega)}$ at the center of the bandgap and the relative amplification bandwidth $\Delta k^\mathrm{a}_{||}$ as a function of the damping factor $\gamma$ for the bandgap shown in (c)--(d). Note that the band structure shown in (b) accommodates the eigenmodes for both TM and TE polarizations. However, we optimize the metasurface for TE-waves in (c)--(d). }
\label{fig:pw}
\end{figure*}

Next, we plot the band structures of the suggested time-varying metasurfaces. We start with a modulation frequency of $\omega_\mathrm{m1}=2\pi\times183$~THz as shown in Fig.~\ref{fig:sw}(b). Since $\frac{\omega_\mathrm{m1}}{2}$ is away from the flat bands, this configuration corresponds to a modulated non-resonant metasurface. Moreover, the dispersion around $\omega=\frac{\omega_\mathrm{m1}}{2}$ is linear (see Fig.~\ref{fig:sw}(b)), indicating that the metasurface is spatially homogenizable near and below $\omega=\frac{\omega_\mathrm{m1}}{2}$ \cite{ramkrishna2021lower, garg2024two}. Therefore, this time-varying metasurface compares in its response to a conventional (space-uniform) PTC with a modulated plasma frequency since the dispersion relation only weakly changes close to the considered $\Gamma$-point (compare Figs.~\ref{Fig:bulk}(b) and (d)).
Figure~\ref{fig:sw}(d) shows the band structure of this non-resonant metasurface. We choose $m=0.01$, corresponding to a relative refractive index change of $\Delta n /n \approx m \chi_0/(1+\chi_0) = 1\% $, close to the maximum attainable value in silicon before the thermal damage occurs~\cite{cowan_optical_2006,polyanskiy_nonlinear_2024}.
We observe the folding of the band structure of the time-invariant case~\cite{Joannopoulos2008photonic,zurita2009reflection1}. 
At the band crossing in the $M -\Gamma$ region, the momentum bandgap appears, as seen in the zoomed region (Fig.~\ref{fig:sw}(e)). 
The calculated relative gap width is very narrow, $\Delta k_{||}= 0.0183\%$. 
We also plot the imaginary part of the frequency $\Im(\omega)$ for a fixed $\Re(\omega)=\frac{\omega_\mathrm{m1}}{2}$ (see Fig.~\ref{fig:sw}(f)). As expected, in the middle of the bandgap, $\Im(\omega)$ reaches the largest absolute value. 
Since the bandgap lies below the light line (see Fig.~\ref{fig:sw}(b)), only surface-wave incident excitations couple to the modes inside it. 

In sharp contrast to the previous case, when the half of the modulation frequency corresponds to the magnetic Mie resonance of the nanospheres (see the horizontal line $0.5{\omega_\mathrm{m2}}=2\pi\times196$~THz in Fig.~\ref{fig:sw}(b)), the bandgap dramatically expands. Keeping the same modulation strength $m=0.01$ for a fair comparison, the relative bandgap size of this resonant metasurface reaches $\Delta k_{||} =6.59\%$, calculated from the band structure in Figs.~\ref{fig:sw}(g)--(i).
Remarkably, the resonance of the metasurface leads to the widening of the bandgap by a factor of about $350$ compared to the non-resonant case (see Supplementary Section 7 for more details on effect of the structural resonances). It should be noted that the correspondence between a time-varying system and two stationary systems with scaled parameters also holds for this metasurface (see Supplementary Fig.~S6). 
Comparing Figs.~\ref{fig:sw}(f) and (i) reveals that the structural resonance not only enhances the bandgap size, but the amplification rate also significantly increases: $\Im{(\omega)}$ is an order of magnitude larger for the resonant case. 

We also study the effects of material losses by considering the nanospheres made from a time-varying dispersive material (see Supplementary Section 8). Depending on the damping factor $\gamma$ of such material, Fig.~\ref{fig:sw}(j) shows the relative amplification bandwidth $\Delta k^\mathrm{a}_{||}$ (defined in the caption) and the imaginary part of the eigenfrequency $\Im{(\omega)}$ at the bandgap center. Regarding this time-varying metasurface, modulation frequency $\omega_{\rm m2}$ is assumed. As $\gamma$ increases, the amplification rate of the eigenmode inside the bandgap $\Im{(\omega)}$ decreases in amplitude. A similar trend is observed for the relative momentum bandwidth where amplification occurs. Nevertheless, amplification remains significant over a wide momentum bandwidth of $\Delta k^\mathrm{a}_{||}> 6\%$  if $\gamma/\omega{\rm r} < 0.05$. Serendipitously, this material loss threshold is significantly above the realistic damping factor $\gamma/\omega_{\rm r} \sim 10^{-5} $ that silicon has at near-infrared wavelengths where the metasurface is designed to operate~\cite{vetterl2000intrinsic}.

While in the above-considered scenarios, the bandgaps occurred below the light line, we optimized next the metasurface to sustain bandgaps for propagating waves (Fig.~\ref{fig:pw}(a)). 
From Fig.~\ref{fig:sw}(b) for a stationary metasurface, we observe a flat band within the light cone at $\omega=\omega_\mathrm{m3}/2=2\pi\times214$~THz around the $\Gamma$-point. Because the modes in the flat band lie within the light cone, they correspond to guided resonance modes \cite{fan2002analysis}.
By modulating the metasurface at $\omega_\mathrm{m3}$, we obtain the band structure shown in Fig.~\ref{fig:pw}(b) ($m=0.06$ in this case). 
Remarkably, despite the small modulation strength, the momentum bandgap is very large and spans over a wide range of incident angles (see Figs.~\ref{fig:pw}(c)--(d)): up to $54^\circ$ in the incidence plane parallel to $M -\Gamma$ and $33^\circ$ parallel to $\Gamma -X$. Due to the almost constant $\Im{(\omega)}$ inside the bandgap (Fig.~\ref{fig:pw}(d)), the waves that couple to the eigenmodes in the bandgap get amplified at nearly the same rates irrespective of the value of $\mathbf{k_{||}}$. 
The bandgap in Fig.~\ref{fig:pw}(b) does not span all possible $\mathbf{k}_{||}$ because of the relatively low-quality factor of magnetic Mie resonances in the nanospheres, as predicted in the LC circuit analysis, and a limited value of $m$. We foresee that by tuning $\omega_{\rm m}/2$ to the frequency of the higher-order Mie resonances of the nanospheres, the bandgap size for the optical metasurface can be further enhanced.
Finally, Fig.~\ref{fig:pw}(e) shows the loss influence on the metasurface operation similarly to the previous example. Here, the influence of losses is even less pronounced due to the larger modulation strength considered in this example.

\section{Discussion} 
Our findings underscore that by exploiting structural resonances in PTC metasurfaces, we can achieve a great enhancement of the momentum bandgap size (350 times wider compared with the same metasurface operating far away from the structural resonances) with the modulation strength as small as 1\%. In principle, a stronger resonance has the potential to decrease the required modulation strength further.
A distinctive feature of our approach is that the achieved momentum bandgap can cover the entire $k$-space, encompassing both free-space propagating modes and surface modes. 
This fundamentally provides new physics compared to the implementations based on bulk media \cite{lustig2018topological,hayran2022homega} that only support propagating modes and non-resonant metasurfaces~\cite{wang2023metasurface} that operate exclusively with surface modes. 
From this perspective, our approach offers novel opportunities for designing more complex photonic time and space-time crystals and for amplifying the spontaneous emission of light from emitters near the structure. 
Indeed, our geometry is practical for the latter application since the emitter does not have to be immersed inside a solid material.
Moreover, the proposed PTCs can be used for designing a perfect lens, a long-standing goal in optics, since the information of an object carried by evanescent modes can be effectively amplified, resulting in a highly-resolved image of the object.
While our PTCs operate in the infrared spectrum, they can be implemented for the visible spectrum with other materials. 
In addition, the shape of the meta-atoms is not restricted to spherical, and they can be deployed over a substrate. We anticipate and encourage experimental endeavors to facilitate the proposed approach. 

\section{Materials and methods}
\subsection{Calculation of eigenmodes of a metasurface with time-varying nanospheres}
Eigenmodes of a scattering structure are self-standing modes that exist without incident excitation. We use the T-matrix method to evaluate the eigenmodes of the time-varying metasurface. For details on the method, see Supplementary Section 6. Combining Eqs.(S28) and (S29), we write
\begin{equation}
\begin{split}
\mathbf{A}^\mathrm{sca}= \left(\mathbf{\hat{U}}-\mathbf{\hat{T}^\mathrm{(s)}(\omega)}\cdot\sum_{\mathbf{R'}\neq\mathbf{0}}\mathbf{\hat{C}}^{(3)}(-\mathbf{R'})\hspace{1pt}\mathrm{e}^{-j\mathbf{k_{{||}}}\cdot\mathbf{R'}}\right)^{-1}\\\cdot\mathbf{\hat{T}^\mathrm{(s)}}(\omega)\cdot\mathbf{A}^\mathrm{inc}\,.\label{eq:Teff2}
\end{split}
\end{equation}
Next, we use $\mathbf{A}^\mathrm{inc}=0$ in Eq.~\eqref{eq:Teff2}. Therefore, we can rewrite Eq.~\eqref{eq:Teff2} as
\begin{eqnarray}
\left(\mathbf{\hat{U}}-\mathbf{\hat{T}^\mathrm{(s)}(\omega)}\cdot\sum_{\mathbf{R'}\neq\mathbf{0}}\mathbf{\hat{C}}^{(3)}(-\mathbf{R'})\hspace{1pt}\mathrm{e}^{-j\mathbf{k_{{||}}}\cdot\mathbf{R'}}\right)\cdot\mathbf{A}^\mathrm{sca}=0\,.\label{eq:Teff3}
\end{eqnarray}
Finally, for Eq.~\eqref{eq:Teff3} to have a non-trivial solution i.e., $\mathbf{A}^\mathrm{sca}\neq0$, we arrive at the condition

\begin{equation}
\label{eq: Tmat_BS}
\left|\mathbf{\hat{U}}-\mathbf{\hat{T}^\mathrm{(s)}(\omega)}\cdot\sum_{\mathbf{R'}\neq\mathbf{0}}\mathbf{\hat{C}}^{(3)}(-\mathbf{R'})\hspace{1pt}\mathrm{e}^{-j\mathbf{k_{{||}}}\cdot\mathbf{R'}}\right|=0\,.
\end{equation}
\noindent
Here, the values of $\omega$ and $\mathbf{k_{||}}$ for which Eq.~\eqref{eq: Tmat_BS} is satisfied correspond to the location of the eigenmodes of the time-varying metasurface. Note that instead of calculating the determinant $D$, we minimize the lowest singular value $S_\mathrm{min}$ of the matrix in Eq.~\eqref{eq: Tmat_BS} to identify the eigenmodes of the system. The metric of the $S_\mathrm{min}$ is advantageous over that of the determinant (see Supplementary Section 9 for more details). Moreover, for plotting the band structures, we limit ourselves only to the dominant dipolar moments in the vector spherical expansion and to the three dominant frequency harmonics (see Supplementary Section 6). This assumption does not modify the response of the metasurface inside the momentum bandgap but only allows the elimination of side-bands in the band structure.

Further, note that for the calculations involving the metasurface made from nanospheres, the radius of the nanospheres $R$ is fixed at $210.6\ \mathrm{nm}$, and the lattice period $a$ is set to $3R$. Moreover, the material susceptibility for the calculations involving dispersionless 
nanospheres is chosen to be $\chi_0=10.68$. Such a value of $\chi_0$ corresponds to silicon, which is nearly lossless and dispersionless in the considered infrared frequency regime~\cite{vetterl2000intrinsic}.

\subsection{Numerical simulations for a time-varying \textit{LC} metasurface}
The simulation results presented in Fig.~\ref{Fig:bigfigure}(c-e),(g) are generated using COMSOL Multiphysics. The metasurface is modeled as an LC parallel circuit with surface current density defined as
\begin{equation}
\mathbf{J}_{\rm s}=\mathbf{J}_{\rm L}+\frac{d[C(t)\mathbf{E}(t)]}{dt},
\end{equation}
where $\mathbf{E}(t)$ is the tangential electric field on the surface. Here, $\mathbf{J}_{\rm L}$ is related to $\mathbf{E}(t)$ by $\mathbf{E}(t)=L\frac{d \mathbf{J}_{\rm L}}{dt}$. This relationship is established via the Boundary Ordinary Differential Equation (ODE) and Differential Algebraic Equation (DAE) module in COMSOL transient simulation.

At $t=0$, a stationary LC surface is excited from the left by a Transverse Magnetic (TM)-polarized surface wave with momentum $k_{||}=5k_{\rm r0}$ (see Fig.~\ref{Fig:bigfigure}(c)) and $k_{||}=10k_{\rm r0}$ (see Fig.~\ref{Fig:bigfigure}(d)). For $t>0$, harmonic modulation of the effective capacitance is initiated, opening the momentum bandgap and resulting in an exponential growth of the surface mode and higher-order frequency harmonics, some of which propagate inside the light cone. This growth is observed in the magnetic-field snapshot at $t = 30 T_{\rm m}$ ($T_{\rm m} = 2 \pi /\omega_{\rm m}$) depicted in the lower panel of Fig.~\ref{Fig:bigfigure}(c).

Temporal modulation keeps the momentum of eigenwaves unchanged but generates backward and forward harmonics with equal amplitudes, creating a standing wave along the horizontal direction~\cite{wang2023metasurface}.

\section*{Acknowledgments}
\begin{acknowledgments}
X.W. and C.R. acknowledge support by the Helmholtz Association via the Helmholtz program “Materials Systems Engineering” (MSE). P.G., A.G.L., and C.R. are part of the Max Planck School of Photonics, supported by the Bundesministerium für Bildung und Forschung, the Max Planck Society, and the Fraunhofer Society. P.G. acknowledges support from the Karlsruhe School of Optics and Photonics (KSOP). P.G. and C.R. acknowledge support by the German Research Foundation within the SFB 1173 (Project-ID No. 258734477). V.A. acknowledges the Research Council of Finland (Project No. 356797), Finnish Foundation for Technology Promotion, and  Research Council of Finland Flagship Programme, Photonics Research and Innovation (PREIN), decision number 346529, Aalto University. 
 X.W. acknowledges the Fundamental Research Funds for the Central Universities, China (Project No. 3072024WD2603) and the Scientific Research Foundation, Harbin Engineering University, China (Project No. 0165400209002).
The authors would like to thank Prof. Sergei Tretyakov for the fruitful discussions on the application of the designed PTCs for constructing a perfect lens. 
\end{acknowledgments}

\section{Author contributions}
X.W. proposed the initial idea of the time-varying resonant structure for enhancing the momentum bandgap, and V.A. expanded this concept to optical metasurfaces to address the challenges in creating optical photonic time crystals. X.W., P.G., M.S.M., A.G.L., and C.R. contributed to the development of the theory of time-varying Lorentz media, while P.G., A.G.L., and C.R. worked on the theory of time-varying optical metasurfaces. P.G. performed the simulations of optical metasurfaces. X.W. and M.S.M. developed the theory and simulation of time-varying LC metasurfaces. X.W., P.G., M.S.M., and V.A. performed data analysis. All authors actively participated in discussions. X.W., P.G., M.S.M., and V.A. prepared the manuscript, with all authors contributing to its review and editing. V.A. and C.R. provided supervision, with C.R. also responsible for funding acquisition.

\section{Competing interests}
The authors declare no competing interests.

\section{Data availability}
The authors declare that the data supporting the findings of this study are available within the paper, in the Supplementary file, and are available from the corresponding authors upon request. 

\section{Code availability}
The codes are available from the corresponding authors upon reasonable request.

\bibliography{references1}

\end{document}